\documentclass[aps,prl,groupedaddresss,superscriptaddress, twocolumn]{revtex4}
\usepackage{graphicx}
\usepackage{amsmath}
\usepackage{bm}
\usepackage{setspace}

\begin{document}
\title{Majorana zero modes by engineering topological kink states in two dimensional electron gas}
\author {Shu-guang Cheng}
\affiliation{Department of Physics, Northwest University, Xi'an 710069, China}
\affiliation{Shaanxi Key Laboratory for Theoretical Physics Frontiers, Xi'an 710069, People's Republic of China}
\author {Jie Liu}
\affiliation{Department of Applied Physics, School of Science, Xi'an Jiaotong University, Xi'an 710049, China}
\author {Haiwen Liu}
\email{haiwen.liu@bnu.edu.cn}
\affiliation{Center for Advanced Quantum Studies, Department of Physics, Beijing Normal University, Beijing 100875, China}
\author {Hua Jiang}
\email{jianghuaphy@suda.edu.cn}
\affiliation{School of Physical Science and Technology, Soochow University, Suzhou 215006, China}
\affiliation{Institute for Advanced Study, Soochow University, Suzhou, 215006, China}
\author {Qing-Feng Sun}
\affiliation{International Center for Quantum Materials, School of Physics, Peking University, Beijing 100871, China}
\affiliation{Collaborative Innovation Center of Quantum Matter, Beijing 100871, China}
\affiliation{CAS Center for Excellence in Topological Quantum Computation, University of Chinese Academy of Sciences, Beijing 100190, China}
\author {X. C. Xie}
\affiliation{International Center for Quantum Materials, School of Physics, Peking University, Beijing 100871, China}
\affiliation{Collaborative Innovation Center of Quantum Matter, Beijing 100871, China}
\affiliation{CAS Center for Excellence in Topological Quantum Computation, University of Chinese Academy of Sciences, Beijing 100190, China}

\begin{abstract}
Majorana zero modes (MZMs)--bearing potential applications for topological quantum computing--are verified in quasi-one-dimensional (1D) Fermion systems, including semiconductor nanowires, magnetic atomic chains, planar Josephson junctions. However, the existence of multi-bands in these systems makes the MZMs fragile to the influence of disorder. Moreover, in practical perspective, the proximity induced superconductivity may be difficult and restricted for 1D systems. Here, we propose a flexible route to realize MZMs through 1D topological kink states by engineering a 2D electron gas with antidot lattices, in which both the aforementioned issues can be avoided owing to the robustness of kink states and the intrinsically attainable superconductivity in high-dimensional systems. The MZMs are verified to be quite robust against disorders and the bending of kink states, and can be conveniently tuned by varying the Rashba spin-orbit coupling strength. Our proposal provides an experimental feasible platform for MZMs with systematic manipulability and assembleability based on the present techniques in 2D electron gas system.
\end{abstract}
\maketitle

Majorana zero modes (MZMs) have potential applications as building blocks for topological quantum computing\cite{Rev01,Rev03}, and been realized in semiconducting nanowires\cite{nanowire1,nanowire1-3,nanowire3,weakCP2,nanowire5}, magnetic atom chains\cite{MagAtom1,MagAtom3}, and planar Josephson junctions\cite{JJ1,JJ2,JJ3,JJ4}, which is scalable and suitable for braiding\cite{Rev02,Braiding}. Recently, the indisputable quantized conductance of MZMs is observed in superconductor coating $\mathrm{InSb}$ nanowires\cite{nanowire5}. However, multi-bands are involved inevitably in all aforementioned quasi-1D systems\cite{Diso}, and the multi-bands overlap with each other under influence of disorder or nanowire bending, which makes the MZMs fragile\cite{Dis1}. Furthermore, the geometry of quasi-1D systems impose harsh constraint to attain relatively high superconducting temperature by proximate effect. Thus, attempts for 1D system hosting single band become natural strategies to overcome these drawbacks. Along this route, there exist proposals based on topological edge states in quantum-spin-Hall insulator and trivial edge states in transition-metal dichalcogenides\cite{3DTI1,3DTI2,MoS2b}. In contrast, little progress has been made in experiments, which may be due to the hardly attainable superconductivity for the former case and the mutability of edge states for the latter. Until recently, MZMs is reported in topologically protected hinge states at the edges of bismuth\cite{MagAtom4}, by virtue of the mutual advantages of the robust 1D topological edge states and the attainable superconductivity in bulk states, which has opened wide apertures to realize MZMs in artificially designing and manufacturing systems with topological edge states.

The topological kink states in graphene systems can manifest as a good candidate for realizing the 1D topological superconductor with MZMs. Previously, such states are realized experimentally\cite{Bilayer1,Bilayer2,Bilayer4}, and reported to be robust against disorder\cite{Class1,Class4}. However, the spin orbit coupling is weak in pristine graphene\cite{weakR}. Such drawback can be overcome in artificial graphene-like system in two-dimensional electron gas (2DEG), which has been constructed in experiment with antidot lattice\cite{Argra0a,Argra3,Argra3a,Argra4,Argra5}. Importantly, recent breakthrough of precisely engineering artificial graphene band in high mobility $\mathrm{GaAs}$ quantum well\cite{Argra3,Argra3a} has pave the way to generating pure 1D topological kink states with tunable Rashba spin orbit coupling (RSOC) and intrinsic superconductivity\cite{RSC1,RSC4,RSC2,RSC3}. Taking advantages of modern semiconductor advantages and the robustness of topological kink states, the  high integration and precise manipulation of MZMs can be gradually achieved, which may bloom the study of artificial graphene in application aspects.

\begin{figure}
\includegraphics[width=\columnwidth, viewport=122 84 616 473, clip]{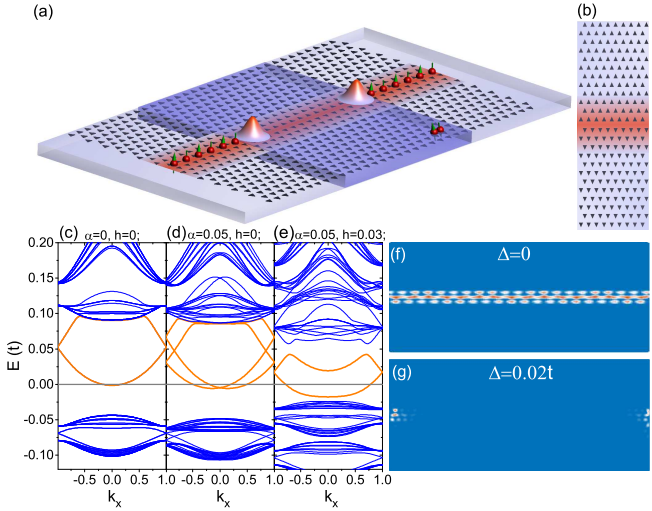}
\caption{Illustration of topological kink states and MZMs in 2DEG. (a) Schematics for constructing MZMs in 2DEG. Triangular antidot lattice with a domain wall (the red path) carries the topological kink states. The shaded blue region refers to the superconductor area, and two red packets at the ends of the superconductor represent the MZMs. The energy bands for 2DEG with antidot lattice with a domain wall (shown in b) are shown in (c-e), with $\alpha$ denoting the Rashba spin-orbit coupling strength and $h_B$ representing the exchanged field. The DOS in the isolated blue region of (a) is shown for without (f) and with (g) superconducting pair $\Delta$.}\label{Fig1}
\end{figure}

In this paper, we propose a practical realization for MZMs based on 2DEG hosting topological kink states, as schematically shown Fig.\ref{Fig1} (a). The 1D topological kink states are formed along the domain wall by engineering asymmetric triangular antidot lattice. Moreover, the topological kink states develop into 1D topological superconductor with MZMs under proper magnetic exchange field, when taking into account the RSOC and (intrinsic or proximity) superconducting effect. Both the quantized tunneling  conductance and the density of states (DOS) locating at the ends of the superconducting domain wall clearly demonstrate the formation of MZMs. We numerically demonstrate that the MZMs are robust against domain wall bending, antidot vacancies and dislocations, originating from topological protected nature of kink states. Our proposal provides an alternative experimentally feasible platform to realize MZMs with topological kink states.\\
\textbf{Results}

\textbf{Topological kink states and MZMs in 2DEG} We start to construct an artificial graphene model \cite{Argra3,Argra3a} in 2DEG with periodic triangular antidot lattice, as displayed in Fig.\ref{Fig1}(b). The Hamiltonian reads:
\begin{equation}\label{EQ1}
H_{2DEG}(\mathbf{k})=-\frac{\hbar^2\bigtriangledown^2}{2m^*}+\alpha_R (\mathbf{\sigma}\times \mathbf{k})_z+h_B\sigma_z+\mu
\end{equation}
in which the first term stands for the kinetic energy and the second term represents the RSOC with strength $\alpha_R$. The third term $h_B$ is for the magnetic exchange field and $\sigma$ the Pauli matrix. The above model is discretized into a square lattice with constant $a$ and nearest coupling $t=\hbar^2/2m^*a^2$. The antidot lattice with $\mathcal{C}_6$ symmetry breaking (as shown in Fig.\ref{Fig1}(b)) is introduced by assuming infinite potentials inside the triangular antidots, and the graphene-like band can be tuned with the size, the shape and the lattice constant of the antidots\cite{Argra7}. The $\mathcal{C}_6$ symmetry breaking strength can be finely tuned in our model, e.g. the orientation of the antidot, and thus the band gap is tunable (see Supplementary Information).

A domain wall is constructed after rotating the triangle antidots by angle $\pi$ in the lower half plane, and the topological kink state locates inside the band gap (see Fig.\ref{Fig1} (c)) due to the mass term change for the upper and lower half plane. Due to strong RSOC in 2DEG, the spin degenerated topological kink states are split as shown in Fig.\ref{Fig1} (d), and an out-of-plane exchange field can further split the bands and leave a single band at the Fermi level (Fig.\ref{Fig1} (e)). The band structures in Fig.\ref{Fig1} (d-e) are similar to the seminal proposal for MZMs in 1D semiconductor nanowires\cite{Hamil3,Hamil4}, thus the topological kink states in 2DEG with superconductivity pairing can act as a good candidate for MZMs.

Taking the superconducting effect into consideration, equation (\ref{EQ1}) is rewritten to the Bogoliubov-de Gennes Hamiltonian:\cite{Hamil1}
\begin{equation}\label{EQ2}
H_{BdG}=\left(
  \begin{array}{cc}
   H_{2DEG}(\mathbf{k}) & i\Delta \sigma_y \\
-i\Delta^* \sigma_y  & -H^*_{2DEG}(-\mathbf{k})\\
  \end{array}
\right)
\end{equation}
with $\Delta$ is the superconducting pair potential. Following the above discretization procedure, we find the topological kink states becomes 1D topological superconductor with MZMs locating at the end of the superconductor, as shown in the schematic DOS in Fig. \ref{Fig1} (a). In the absence of superconductor the kink states are well restricted around the domain Fig.\ref{Fig1} (g); When the superconductor is present, peaks of DOS are concentrated at the edges, demonstrating the formation of MZMs.

\begin{figure}
\includegraphics[width=\columnwidth, viewport=48 75 634 524, clip]{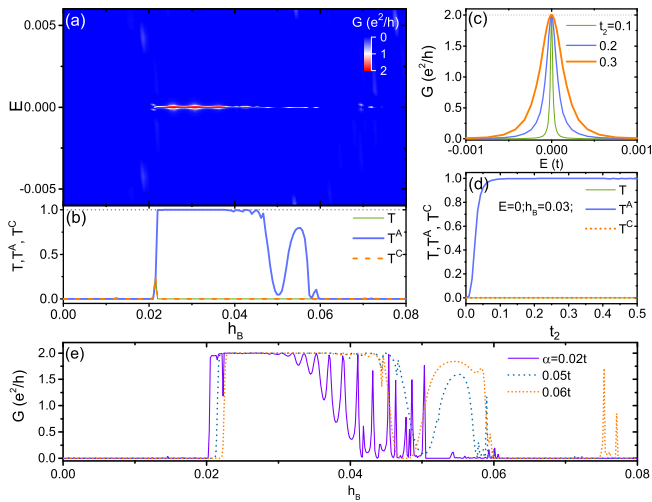}
\caption{Conductance characterized MZMs. (a) Fermi level $E$ and exchange field $h_B$ dependence of conductance $G(E,h_B)$. (b) $G$ as a function of $h_B$ for $E=0$. (c) $G$ as a function of $E$ for different weak coupling strength $t_2$ ($h_B=0.03$). (d) Normal transmission $T$, lateral Andreev reflection $T^A$ and crossed Andreev reflection $T^C$ coefficients versus $t_2$ for $E=0$ and $h_B=0.03t$. (e) The zero-bias conductance $G$ as a function of exchange $h_B$ for different RSOC strength $\alpha$.
}\label{Fig2}
\end{figure}

For the purpose of comparison to experiments, lattice constant $a=5 \mathrm{nm}$ is used thus the typical size of a single antidot is about $30 \mathrm{nm}$\cite{Argra3,Argra3a}. We set the width/length of the superconductor area (see Fig.\ref{Fig1} (a)) as $0.85\mu m$/4$\mu m$ which includes $20$/$40$ antidot supercells in the transverse/longitudinal direction. The whole length of the domain wall is $8\mu m$. Here $t=\hbar^2/2m^*a^2$ is about $31\mathrm{meV}$ with $m^*=0.05m_e$. Also $\Delta=0.02t$ and $\alpha=\alpha_R/2a=0.05t$ is adopted unless specified. So the corresponding superconductor gap is $\Delta=0.02t\sim0.62\mathrm{meV}$ and the RSOC strength is $\alpha_R\sim0.15\mathrm{eV {\AA}}$. In experiment to highlight the role of MZMs assisted electron transport by Andreev reflection, either a potential barrier or a quantum dot is introduced for weak coupling\cite{nanowire1,weakCP2,nanowire5}. In the present model the superconductor area is weakly coupled at two ends to the topological kink states with a parameter $t_2$ ($t_2=0.2t$ is used unless specified).


\textbf{Transport characteristics of MZMs}
Transport measurements with quantized zero-bias conductance peak ($2e^2/h$) can strongly support the MZMs\cite{nanowire5}. In our proposed system, the Fermi level $E$ dependent conductance $G(E,h_B)$ at zero temperature is obtained based on Green's function method and Landau-B\"{u}ttiker formula. Here $G(E,h_B)=(T+2T^{A}+T^{C})e^2/h$ is contributed by normal transmission $T$, lateral Andreev reflection $T^A$ and crossed Andreev reflection $T^C$, respectively. The main results are shown in Fig.\ref{Fig2} (a). The quantized conductance is found at $E=0$ when $h_B>\sqrt{\Delta^2+\mu_0^2}\sim0.021t$. Correspondingly, the band gap of the system closes and reopen at $h_B=0.021t$ (see S.2 in the Supplementary). The detailed conductance are plotted in Fig.\ref{Fig2} (b) and (c). At zero $E$, $T^A$ tunes from zero to $1$ when $h_B>0.021t$ and $T^A$ oscillates to zero when $h_B$ is larger\cite{weakCP,weakCP2,Osci1}. The normal transmission and crossed Andreev reflection, highly suppressed by the weak coupling, only shown a tiny peak when the quantum phase transition happens. Such features demonstrate the existence of MZMs. According to the criterion for 1D superconductor, the superconductor area is under topological superconductor phase when $h_B>\sqrt{\Delta^2+\mu_0^2}$. Moreover, under very large $h_B$, multi-bands pass through the Fermi level and the system returns topological trivial states (see the band evolution in Supplementary Information). Thus, within the $h_B$ interval $[0.021t,  ~0.062t]$, MZMs dominate the electron transport. The conductance is mainly contributed by the lateral Andreev reflection and quantized $T^{A}$ is observed\cite{zero}. In Fig.\ref{Fig2} (c), narrow peaks of $G$-$E$ are shown with different $t_2$. Mind that near $E=0$, $G$ is only contributed by $T^A$. When the coupling is strengthened, the peak of $G$ is expanded and the peak value $G=2e^2/h$ still hold. $G$-$E$ relation is in good agreement with zero-bias peak theory of MZMs. Previously, we use $t_2=0.2t$ to underline the electron transport via MZMs. In Fig.\ref{Fig2} (d), $T$, $T^A$ and $T^C$ versus $t_2$ relations are displayed for $E=0$ and $h_B=0.03t$. Starting from a tiny $t_2$ ($\sim 0.06$), $T^A$ is nearly $1$ and it lasts for larger $t_2$\cite{nanowire5}, meanwhile both $T$ and $T^C$ are zero. The quantized $G$ with weak coupling at zero energy reverifies the formation of MZMs in our system.

Compared to 1D semiconductor nanowires, the RSOC in 2DEG can be adjusted more easily and precisely. It advantages from the experiment manipulation of MZMs\cite{RSC2}. The $G$-$h_B$ relation are shown in Fig.\ref{Fig2} (e) with different $\alpha$. For all $\alpha$ values, the critical $h_B$ for topological superconductor regime is nearly unaltered. When $\alpha$ is smaller (e.g. $\alpha=0.02t$), the quantized $G$ plateau shows in a small window. $G$ oscillates and decreases when $h_B$ becomes larger\cite{weakCP2,weakCP}. By enhancing $\alpha$, the quantized $G$ plateau is wider and more stable. It means stronger RSOC is favored for the formation of MZMs. In experiments, the RSOC strength can be tuned continuously by an external electric field\cite{RSC4,RSC2}.

\begin{figure}
\includegraphics[width=8.5cm, viewport=173 263 484 506, clip]{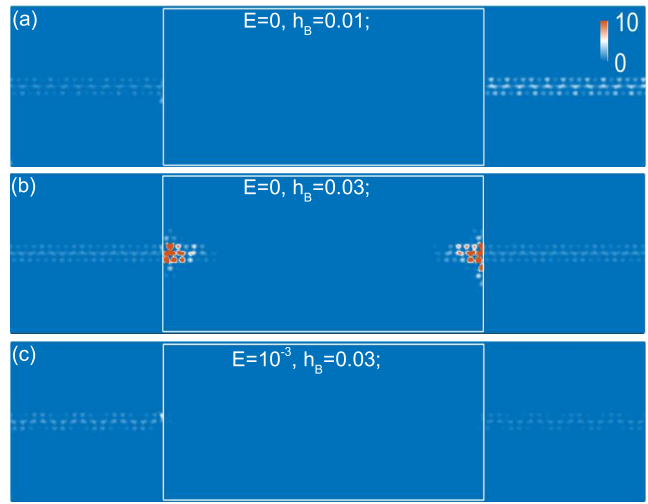}
\caption{DOS properties of topological kink states and MZMs. (a-c) The DOS of central region under various exchange field $h_B$ with energy $E=0$. The white rectangular signifies the superconductor area. In (a), only the topological kink states are observed. In (b), both MZMs and topological kink states exist. Nevertheless the kink states are indistinct because the magnitude of DOS for MZMs is much larger. (c) When $E$ is shifted lightly from zero energy, the MZMs signal is subtle. 
}\label{Fig3}
\end{figure}

\textbf{The DOS characteristics of MZMs}
In experiment, the MZMs can be characterized by detecting the DOS at the ends of the nanowire with scanning tunneling microscope techniques\cite{MagAtom1,MagAtom3}. Here to further examine the properties of MZMs, the features of DOS is investigated in Fig.\ref{Fig3}. When $h_B=0.01t$ and $E=0$, the kink states are under trivial superconducting region. The DOS concentrates along the domain wall in the normal region. Within  the superconductor area, DOS is zero because of the superconductor gap, as shown in Fig.\ref{Fig3} (a). When the exchange field is increased, the band gap is closed and reopened (see Supplementary Information). The system undertakes a phase transition into topological superconducting region. Two sharp DOS peaks emerge at the ends of superconducting topological kink states in Fig.\ref{Fig3} (b). By further enhancing the exchange field, the band gap is closed and reopened again and system subjects into a topological trivial phase. 
When $E$ moves away from the zero value, the effect of MZMs is inactive, as shown in Fig.\ref{Fig3} (c). The peaks of DOS in the superconducting area is strongly suppressed and meanwhile the conductance is nearly zero (see Fig.\ref{Fig2} (c)). The spatial size of the MZMs DOS peaks in Fig.\ref{Fig3} (b) is about $100\mathrm{nm}\times100\mathrm{nm}$, so it is easily for detecting with scanning tunneling microscope techniques in experiments\cite{MagAtom1,MagAtom3}.

\begin{figure}
\includegraphics[width=\columnwidth, viewport=187 47 538 561, clip]{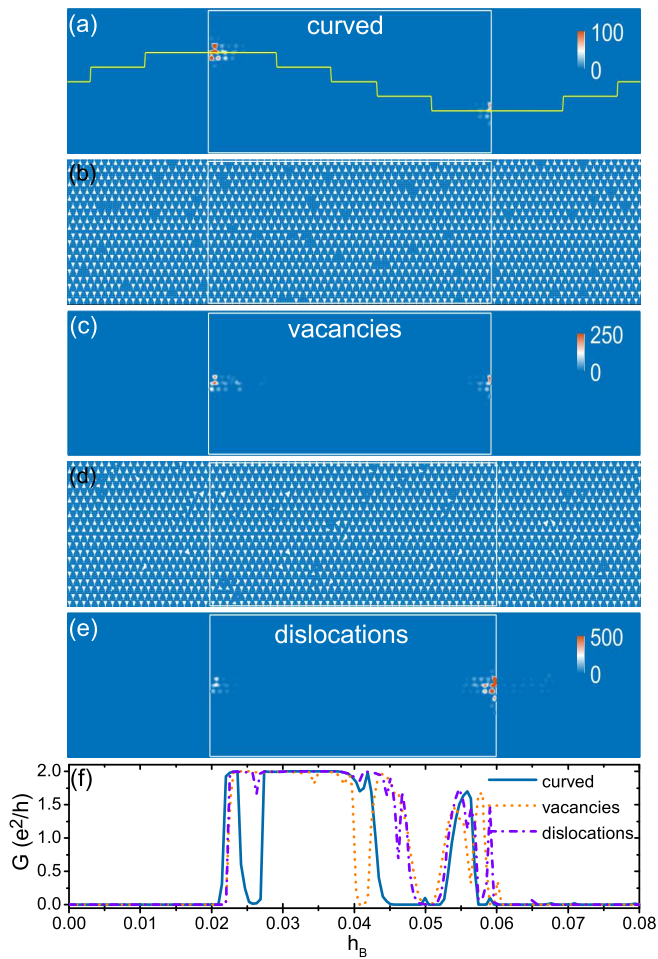}
\caption{Stability of MZMs. (a, c, e) DOS of the central region under three cases: curved domain wall (the yellow path in a), antidot vacancies (b) and antidot dislocations (d). In all the cases, the MZMs are stable. (f) Zero-bias conductance $G$ as a function of exchange field $h_B$ for different flawed antidot lattices (a, c, e). Quantized $G$ is observed in the topological superconducting region in all cases.
}\label{Fig4}
\end{figure}

For the purpose of Majorana braiding,\cite{Braiding} stably moving of MZMs is favored. In host systems from trivial states, the MZMs are fragile to the bending of nanowire or disorders which brings the mixture of subgap states.\cite{Dis1,Dis3} During the fabrication process for our proposed model, the domain wall bending, antidot vacancy, and antidot dislocation are unavoidable. Thus, systematic investigations are needed to uncover the effect of these three types disorder on the formation of MZMs. For the first case, the curved domain wall (the yellow path in Fig.\ref{Fig4} (a)) is determined by the pattern of antidot lattices. In the non-superconductor area, DOS is large along the curved domain (suppressed by the sharp DOS peaks of MZMs in Fig.\ref{Fig4} (a), see Supplementary S.3(d)), indicating the formation of topological kink states. At the intersections between kink states and the superconductor edges, sharp DOS peaks of MZMs are found. Thus the MZMs from topological kink states are robust against the domain wall bending. For the second case, random antidot vacancies are considered (Fig.\ref{Fig4} (b), with a probability $3\%$). Compared with a flawless model in Fig.\ref{Fig3} (b), the DOS peaks of MZMs, still well located at the edges of superconductor, are nearly unaffected. For the last case, the triangular shaped antidots rotate randomly as displayed in Fig.\ref{Fig4} (d). Similar characteristics of DOS peaks are shown, signifying the robustness of MZMs under such a kind of disorder. Furthermore, in all three cases $G$ shows stable quantized zero-bias conductance peak in the topological superconductor region, as displayed in Fig.\ref{Fig4} (f). These results clearly demonstrate that the zero energy MZMs are formed at the ends of topological kink states and robust against the kink states bending, antidot vacancies or random dislocations, protected by the topological nature of nontrivial kink states.

\textbf{Discussion}

Here, a platform based on superconducting 1D topological kink states in 2DEG is proposed to generate MZMs. In 2DEG, the superconductor can be induced intrinsically or through 2D proximity effect. For instance, at the interface of $\mathrm{LaAlO}_3/\mathrm{SrTiO}_3$\cite{RSC3} or 2D GaAs heterostructures\cite{RSC4}, both the superconductor and RSOC are observed. Thus, MZMs based on topological kink states are experimentally feasible, which can be tuned by varying the RSOC strength and the superconductor gap. Moreover, due to the topological nature of kink states, MZMs in such a platform are robust against weak perturbations, e.g. the aforementioned various types of disorder. On the other hand, in addition to the the stability of MZMs, the key road towards to topological quantum computation is a scalable structure\cite{Karzig2017}. As shown in Figure 4(a), the MZMs remain stable against various curved domain wall structure, which means one can easily construct various 2D network structures based on the component unit of kink state\cite{wile2018}. Indeed, recently a square lattice structure have been experimentally constructed through engineer kink states\cite{Li2018}. Thus, engineering kink states may provide advantages for constructing network structures to progressively realizing, manipulating, and non-abelian braiding of MZMs.

\textbf{Methods}

\textbf{Constructing antidot lattice with a tight-binding approach}
We developed a method based on the tight-binding model to dealing with antidot lattice in 2DEG. Equation (\ref{EQ2}) for a 2DEG ribbon is discretized to a square lattice. The antidot triangular lattice is introduced by periodical hard-wall potentials and then the sites inside the antidot area are disconnected. Consequently, the area in 2DEG ribbon allowing electron transport bears a hexagonal lattice-acting as an artificial graphene. Compared with the planewave expansion method\cite{Argra0a}, the present method possesses obvious advantages. i) It is suitable for simulation of actual antidot lattice which do not necessarily has a perfect period (e.g. defects) or even for any geometry of antidots formation. ii) Instead of an infinite system, the treatment applies to a ribbon, a ribbon with domain walls or a sample of any size or shape. iii) In combination with lattice Green's function calculations, both conductance, DOS and the valley resolved transmission coefficients are accessible under various complex situations.

\textbf{Transport and DOS calculation of tight-binding model} The electronic transport properties are calculated by the Green's function method. Using the discretized Hamiltonian, the retarded Green's function is obtained by  $G^r(E)=(E-H_{cen}-\Sigma^r_L-\Sigma^r_R)^{-1}$ with $H_{cen}$ the Hamiltonian of the central region (the lattice part in Fig.\ref{Fig1}(a)). The self-energy $\Sigma^r_{L/R}$ refers to the coupling of the normal 2DEG terminals. The transmission coefficients are calculated by $T(E)=\mathrm{Tr}[\Gamma^L_{11}G^r_{11}\Gamma^R_{11}G^a_{11}]$, $T^A(E)=\mathrm{Tr}[\Gamma^L_{11}G^r_{12}\Gamma^L_{22}G^a_{21}]$ and $T^C(E)=\mathrm{Tr}[\Gamma^L_{11}G^r_{12}\Gamma^R_{22}G^a_{21}]$. Here $1(2)$ refers to the electron(hole) index of the matrix in the Numbu representation. Energy and sites dependent of DOS is calculated by $\rho_\mathbf{i}(E)=-\mathrm{Im}[G^r(\mathbf{i})]/\pi$.\\

\textbf{Acknowledgments}\\ This work was supported by NSFC under Grants Nos. 11674264, 11534001, 11574007, 11674028 and 11822407,  National Key R and D Program of China (2017YFA0303301), NBRP of China (2015CB921102) and the Key Research Program of the Chinese Academy of Sciences (Grant No. XDPB08-4).\\
\textbf{Author contributions}\\
H.J. and  H.W.L. initiate the idea from a discussion. S.G.C and H.J. performed the calculation with the suggestions from H.W.L. and J.L. Q.F.S. and X.C.X. supervised the project. All authors were involved in analyzing and writing of the manuscript.\\
\textbf{Competing financial interests}: The authors declare no competing financial interests.\\

\pagebreak
\newpage

\setcounter{equation}{0}
\setcounter{figure}{0}
\setcounter{page}{1}

\renewcommand{\figurename}{S.}

\begin{widetext}

\begin{center}
\textbf{Supplementary Information for: Majorana zero modes by engineering topological kink states in two dimensional electron gas}
\end{center}

\begin{center}
Shu-guang Cheng$^{1,2}$, Jie Liu$^3$, Haiwen Liu$^{4,*}$, Hua Jiang$^{5,6,\dag}$, Qing-Feng Sun$^{7,8,9}$, and X. C. Xie$^{7,8,9}$
\end{center}

\begin{center}
$^1$~Department of Physics, Northwest University, Xi'an 710069, China

$^2$~Shaanxi Key Laboratory for Theoretical Physics Frontiers, Xi'an 710069, People's Republic of China

$^3$~Department of Applied Physics, School of Science, Xi'an Jiaotong University, Xi'an 710049, China

$^4$~Center for Advanced Quantum Studies, Department of Physics, Beijing Normal University, Beijing 100875, China

$^5$~School of Physical Science and Technology,  Soochow University, Suzhou 215006, China

$^6$~Institute for Advanced Study, Soochow University, Suzhou 215006, China

$^7$~International Center for Quantum Materials, Peking University, Beijing 100871, China

$^8$~Beijing Academy of Quantum Information Sciences, Beijing 100193, China

$^9$~CAS Center for Excellence in Topological Quantum Computation, University of Chinese Academy of Sciences, Beijing 100190, China

\end{center}

\subsection{Artificial graphene band in 2DEG}
To establish an antidot lattice, the tight-binding model for 2DEG is adopted. Discretize equation (1) in the main text into a tight-binding model in a square lattice, we have
\begin{eqnarray}\label{EQS1}
H_{2DEG}&=&\sum_{\mathbf{i} \sigma} c^{\dagger}_{\mathbf{i}\sigma} (\varepsilon_{\mathbf{i}}+h\sigma_z+\mu) c_{\mathbf{i}\sigma}-t\sum_{ \langle \mathbf{ij} \rangle \sigma} (c^{\dagger}_{\mathbf{i}\sigma} c_{\mathbf{j}\sigma}+h.c.) \nonumber \\ &-&\alpha_R\sum_{\langle \mathbf{i} \rangle \sigma\sigma'}[c^{\dagger}_{\mathbf{i}+\delta \mathbf{y}, \sigma}(i\sigma_x)_{\sigma\sigma'}c_{\mathbf{i}\sigma'}
-c^{\dagger}_{\mathbf{i}+\delta \mathbf{x},\sigma} (i\sigma_y)_{\sigma\sigma'}c_{\mathbf{i}\sigma'}\nonumber \\ &+&h.c.].
\end{eqnarray}
with $c^{\dagger}_{\mathbf{i}\sigma}$ the creation operator on site ${\mathbf{i}}$.

The first term is for the on-site energy, the exchange field term and the potential shift. The second and third terms are for nearest neighbor site hopping and RSOC, respectively. To simulate an antidot lattice, the sites in the antidot area are disconnected from all sites around. Thus in the numerical calculation, we are dealing with a supercell as displayed in S.\ref{S1} (a).

Following Bloch's theorem, band structure can be finely tuned by engineered periodical potential lattices. In 2DEG, a triangular antidot lattice leaving the electrons subjecting to a hexagonal lattice, as shown in S.\ref{S1}. In the discretied square lattice, we choose a rectangular region with size $l\times\sqrt{3}l$ (see S.\ref{S1} (a)). In the numerical calculation $\sqrt{3}l$ is an irrational number, thus we replace it with an integer (e.g. $10\times17$ sites). Two circular shaped antidots are introduced along the diagonal line. Their positions are $(l/4, \sqrt{3}l/4)$ and $(3l/4, 3\sqrt{3}l/4)$, respectively. Such supercells combine give to a triangular array acting as a triangle-lattice with constant $l$ (see S.\ref{S1} (b)). The rest part allowing the transport of electrons subjects to a hexagonal lattice. Alternative, one may also choose a square supercell but set the transverse (longitudinal) nearest site hopping as $t$ ($t/\sqrt{3}$). By constructing the antidots at $(l/4, l/4)$ and $(3l/4, 3l/4)$, similar results will be obtained.

\begin{figure}
\includegraphics[width=10cm, viewport=20 144 735 470, clip]{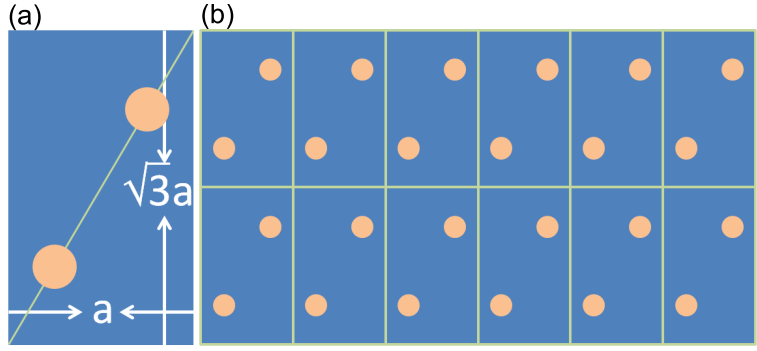}
\caption{(a) A rectangular shaped supercell of antidot lattice in 2DEG. The ratio between the width and length of a single supercell is $\sqrt{3}$. Along the diagonal line, two circular shaped antidots are located at $1/4$ and $3/4$ of the line from either side. (b) The combination of the supercells gives a triangular antidot lattice which forbids the electrons. The rest part forms a hexagonal lattice of electrons in 2DEG, which establishes the graphene-like band in 2DEG.
}\label{S1}
\end{figure}

\begin{figure}
\includegraphics[width=10cm, viewport=54 18 695 540, clip]{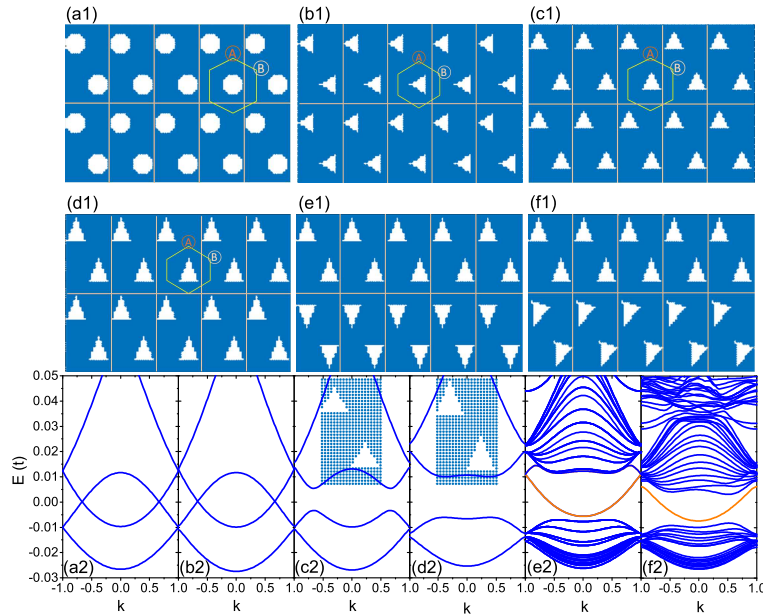}
\caption{(a1-f1) The typical patterns of antidot lattice in 2DEG. (a2-f2) The band structure of corresponding antidot patterns in (a1-f1). Inducing triangular antidots in a square lattice is highlighted in insets of (c2) and (d2). In (f1), the bottom antidot lattice rotates a $0.1\pi$ compared with (e1). Bands in (a1-d1) are for 2D and (e1-f1) are for a ribbon with ten supercells in the transverse direction.
}\label{S2}
\end{figure}

\subsection{Constructing topological kink states with antidot lattice in 2DEG}
If the antidots are of circular shaped as in S.\ref{S2} (a1), the periodical potential bears a $\mathcal{C}_6$ symmetry and the two sub-lattices are equivalent. The graphene-like band appears with zero band gap (S.\ref{S2} (a2)). The calculation reproduce the result in Ref.\cite{Sr1,Sr2,Sr3}. If the antidots are of the triangle shape as in S.\ref{S2} (b1-d1), the $\mathcal{C}_6$ symmetry falls into $C_3$ symmetry. In this case, the AB sublattices can be equivalent (see S.\ref{S2} (b1)). The zero gap graphene-like band shows as well. However, when the triangular shaped antidot rotates a $\pi/2$ in S.\ref{S2} (c1), the area for sublattice $A$ is smaller than that of sublattice $B$, the AB sublattices are no longer equivalent. A band gap appears as displayed in S.\ref{S2} (c2). The sublattice asymmetry is enhanced as the antidot becomes acute (S.\ref{S2} (d2)). Consequently, the band gap becomes larger. By introducing a domain wall in the lattice between which the triangular shaped antidot rotates an angle $\pi$ (S.\ref{S2} (e1)), the topological kink states appear, as displayed in S.\ref{S2} (e2). Here the trivial edged states located at the edges of the ribbon are not plot for clarity (also in the Fig.1 (c-e) in the manuscript and S.\ref{S3}) because the trivial states are fragile to disorder or edge deformation. The topological kink states are stable even the rotation deviates from $\pi$ slightly (see S.\ref{S2} (f1) and (f2) for angle deviation $0.1\pi$). A single band for ideal 1D topological channel across the Fermi level guarantees the possibility for MZMs when taking RSOC, exchange field term and superconductor into consideration.

\begin{figure}
\includegraphics[width=10cm, viewport=60 80 735 515, clip]{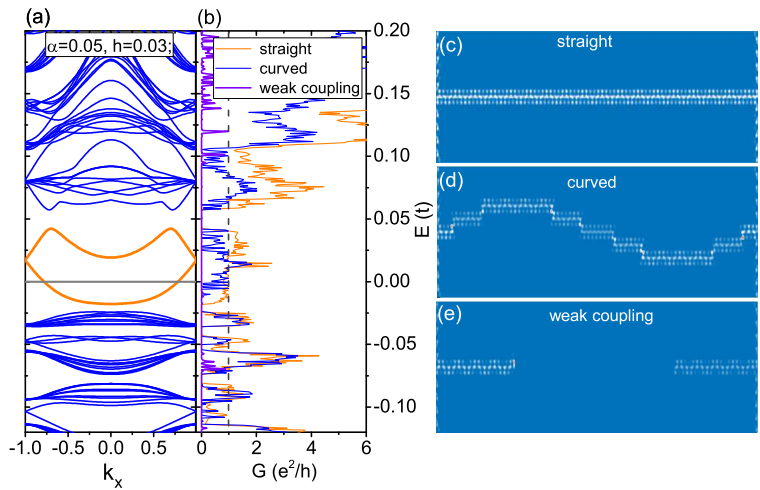}
\caption{(a) Band structure of topological kink states from Fig.1 (e) in the main text. (b) The transport features of topological kink states in three cases: a straight domain wall, a curved domain wall and a straight domain with weak coupling. (c-e) The DOS of the ribbon for the above three cases.
}\label{S4}
\end{figure}

\begin{figure}
\includegraphics[width=10cm, viewport=11 98 680 515, clip]{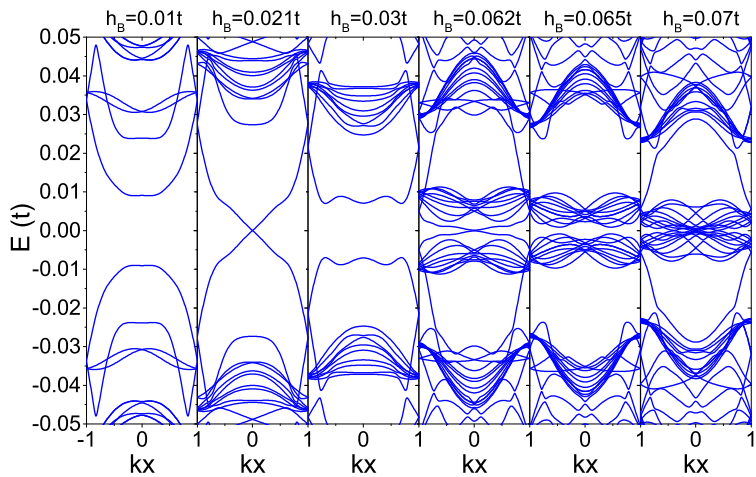}
\caption{Band structure of topological kink states at different exchanged field $h_B$. The sample size are the same to Fig.1 (d) in the main text. Other parameters are RSOC $\alpha=0.05t$ and superconductor pair potential $\Delta=0.02t$.
}\label{S3}
\end{figure}

\subsection{Transport properties of topological kink states}
Now we study the transport properties of topological kink states when superconductor is absent (the same setup as in Fig.1 (a)). The corresponding band structure form Fig.1 (e) is redisplayed in S.\ref{S4} (a) for clarity. Three cases are investigated: a straight domain wall, a curved domain wall and a straight domain wall with weak coupling $t_2$ (the same as the main text). For a straight domain wall, the conductance $G$ is nearly quantized when the Fermi level crosses the single topological band (see S.\ref{S4} (b)). $G$ is suppressed when the domain wall bends. Finally, for weak coupling ($t_2=0.2t$), $G$ is nearly zero within the gap region. DOS features, calculated from the Green's function method,\cite{Sr4} are displayed in S.\ref{S4} (c-e). For both a straight domain wall and a curved domain wall, DOS is well restricted along the domain wall. It means the topological kink state survives when the domain wall bends. When the side-coupling is weak, DOS in the center of the ribbon is nearly zero. So the normal transmission of the topological kink state is strongly suppressed by weak coupling or barrier potentials. For comparison, in the main text the Andreev reflection is almost unaffected by the weak coupling which show evidence for MZMs.

\subsection{Exchange field dependent of band evolution of topological kink states}
Furthermore, we take superconductor into consideration, Eq.\ref{EQS1} is rewritten in the Bogoliubov-de Gennes representation with the non-diagonal pair potential terms $\pm i \Delta \sigma_y$. The band structures of the superconductor ribbon are shown in S.\ref{S3} for different exchange field $h_B$. At $h_B=0.01t$, a trivial superconductor gap appears as displayed in S.\ref{S3}. At $h_B=0.021t$, the gap closes and the quantum phase transition happens. The system tunes into topological superconductor region. When $h_B=0.03t$, the gap reopens and the system is under the topological superconductor region. In the main text, we mainly concentrate on this region $h_B\in[0.021t, 0.062t]$. The gap close again when $h_B=0.062t$ and the system transmits into trivial superconductor. For even larger $h_B$, the superconductor is damaged by the exchange field. The band structures are in good agreement with the transport results in the main text.

\end{widetext}

\end{document}